Charmonium-like structures with the open-charm and the open-strange channels.

Gerasyuta S.M.[1,2], Kochkin V.I.[1]

1. Department of Theoretical Physics, St. Petersburg State University, 198904, St. Petersburg, Russia.
2. Department of Physics, LTA, 194021, St. Petersburg, Russia,
   E-mail: gerasyuta@sg6488.spb.edu

Abstract

The relativistic four-quark equations with the open-charm and the open-strange are found in the framework of coupled-channel formalism. The dynamical mixing of the meson-meson states with the four-quark states is considered. The four-quark amplitudes including the quarks of four flavors (u, d, s, c) are constructed. The poles of these amplitudes determine the masses of tetraquarks. The mass values of the tetraquarks with the spin-parity $J^P = 1^-, 2^-$ are calculated.

I. Introduction.

QCD, the dynamically theory of strong interactions, appears to predict the existence of hadrons beyond those in the simple quark model, namely glueballs, hybrids and $q\bar{q}q\bar{q}$ states. Confirmation of such states would give information of the role of «dynamical» color in the low energy QCD. The spectroscopy of low-mass states can be accounted by QCD-inspired models [1-9].

Phenomenological models are expected to provide a scheme for taking into account vacuum effects and other effects of a non-perturbative character in a zero-order approximation. Investigation revealed that the spectroscopy of light hadrons is one of the main sources of information about non-perturbative QCD effects. The remarkable progress of the experimental side has opened up new challenges in the theoretical understanding of heavy flavor hadrons. The observation of $X(3872)$ resonance [10] has been confirmed by *CDF* [11], *D0* [12] and *BaBar* [13] Collaborations. Belle Collaboration observed the $X(3940)$ in double-charmonium production in the reaction $e^+e^- \to J/\psi + X$ [14]. The state, designated as $X(4160)$, was reported by the Belle Collaboration in Ref. 15. The fact that the newly bound states do not fit quark model calculations has opened the discussion about the structure of such states. Maiani et al. advocate a tetraquark explanation for the $X(3872)$ [16, 17]. On the other hand, the mass of $X(3872)$ is very close to the threshold of $D^*D$ and, therefore, it can be interpreted as molecular state. In our paper [18] the dynamical mixing between the meson-meson states and the four-quark states is considered. Taking the $X(3872)$ and $X(3940)$ as input we predicted the masses and the widths of S-wave tetraquarks with the open and hidden charm [19, 20].

Flavor-exotic tetraquark mesons have recently been observed in the heavy-quark pair sectors of QCD, including two isospin multiplets in the $b\bar{b}$ sector, $Z_b(10610)$ and $Z_b(10650)$, and one isospin multiplet in the $c\bar{c}$ sector, $Z_c(3900)$ [21-25].

After this introduction, we obtain the relativistic four-particle equations, which describe the interaction of heavy and light quarks (Sec. II). Section III is devoted to a calculation of the masses of the low-lying tetraquarks, which contain the u, d, s, c quarks (Table I and II). In the conclusion the considered model is discussed.

II. Four-quark amplitudes for the negative parity tetraquarks.

We derive the relativistic four-quark equations in the framework of the dispersion relation technique. We use only planar diagrams; the other diagrams due to the rules of $1/N_c$ expansion [26-28] are neglected. The current generates a four-quark system. Their successive pair interactions lead to the diagrams shown in Fig. 1. The correct equations for the amplitude are obtained by taking into account all possible subamplitudes. It corresponds to the division of complete system into subsystems with smaller number of particles. Then one should represent a four-particle amplitude as a sum of six subamplitudes:

$$A = A_{12} + A_{13} + A_{14} + A_{23} + A_{24} + A_{34}. \tag{1}$$

This defines the division of the diagrams into groups according to the certain pair interaction of particles. The total amplitude can be represented graphically as a sum of diagrams. We need to consider only one group of diagrams and the amplitude corresponding to them, for example $A_{12}$. The derivation of the relativistic generalization of the Faddeev-Yakubovsky approach for the tetraquark is considered. We shall construct the four-quark amplitude of $c\bar{s}u\bar{u}$ meson. The set of diagrams associated with the amplitude $A_{12}$ can further be broken down into groups corresponding to subamplitudes: $A_1(s,s_{12},s_{34})$, $A_2(s,s_{23},s_{14})$, $A_3(s,s_{12},s_{34})$, $A_4(s,s_{23},s_{14})$, $A_5(s,s_{23},s_{123})$, $A_6(s,s_{14},s_{134})$, $A_7(s,s_{34},s_{234})$, $A_8(s,s_{12},s_{123})$ (Fig. 1), if we consider the tetraquark with $J^P = 2^-$. Here $s_{ik}$ is the two-particle subenergy squared, $s_{ijk}$ corresponds to the energy squared of particles $i$, $j$, $k$ and $s$ is the system total energy squared.

The system of graphical equations is determined by the subamplitudes using the self-consistent method. The coefficients are determined by the permutation of quarks [29, 30].

In order to represent the subamplitudes $A_1(s,s_{12},s_{34})$, $A_2(s,s_{23},s_{14})$, $A_3(s,s_{12},s_{34})$, $A_4(s,s_{23},s_{14})$, $A_5(s,s_{23},s_{123})$, $A_6(s,s_{14},s_{134})$, $A_7(s,s_{34},s_{234})$, $A_8(s,s_{12},s_{123})$ in the form of a dispersion relation it is necessary to define the amplitudes of quark-antiquark interaction $a_n(s_{ik})$. The pair quarks amplitudes $q\bar{q} \to q\bar{q}$ are calculated in the framework of the dispersion

N/D method with the input four-fermion interaction [31, 32] with the quantum numbers of the gluon [33]. The regularization of the dispersion integral for the D-function is carried out with the cutoff parameters $\Lambda$.

The four-quark interaction is considered as an input:

$$g_v(\bar{q}\lambda I_f \gamma_\mu q)^2 + 2g_v^{(s)}(\bar{q}\lambda\gamma_\mu I_f q)(\bar{s}\lambda\gamma_\mu s) + g_v^{(ss)}(\bar{s}\lambda\gamma_\mu s)^2 + \\ + 2g_v^{(c)}(\bar{q}\lambda\gamma_\mu I_f q)(\bar{c}\lambda\gamma_\mu c) + 2g_v^{(cs)}(\bar{s}\lambda\gamma_\mu s)(\bar{c}\lambda\gamma_\mu c) \quad (2)$$

Here $I_f$ is the unity matrix in the flavor space (u, d), $\lambda$ are the color Gell-Mann matrices. In order to avoid additional violation parameters we introduce the scale shift of the dimensional parameters [33]:

$$g = \frac{m^2}{\pi^2}g_v = \frac{(m+m_s)^2}{4\pi^2}g_v^{(s)} = \frac{m_s^2}{\pi^2}g_v^{(ss)} = \frac{(m+m_c)^2}{4\pi^2}g_v^{(c)} = \frac{(m_s+m_c)^2}{4\pi^2}g_v^{(cs)}, \quad (3)$$

$$\Lambda = \frac{4\Lambda(ik)}{(m_i+m_k)^2}. \quad (4)$$

Here $m_i$ and $m_k$ are the quark masses in the intermediate state of the quark loop. Dimensionless parameters g and $\Lambda$ are supposed to be constants which are independent of the quark interaction type. The applicability of Eq. (2) is verified by the success of De Rujula-Georgi-Glashow quark model [34], where only the short-range part of Breit potential connected with the gluon exchange is responsible for the mass splitting in hadron multiplets.

We use the results of our relativistic quark model [33] and write down the pair quark amplitudes in the form:

$$a_n(s_{ik}) = \frac{G_n^2(s_{ik})}{1 - B_n(s_{ik})}, \quad (5)$$

$$B_n(s_{ik}) = \int_{(m_i+m_k)^2}^{(m_i+m_k)^2 \Lambda/4} \frac{ds'_{ik}}{\pi} \frac{\rho_n(s'_{ik})G_n^2(s'_{ik})}{s'_{ik} - s_{ik}}. \quad (6)$$

Here $G_n(s_{ik})$ are the quark-antiquark vertex functions (Table III). The vertex functions are determined by the contribution of the crossing channels. The vertex functions satisfy the Fierz relations. All of these vertex functions are generated from $g_v$, $g_v^{(s)}$, $g_v^{(ss)}$, $g_v^{(c)}$ and $g_v^{(cs)}$.

$B_n(s_{ik})$ and $\rho_n(s_{ik})$ are the Chew-Mandelstam functions with cutoff $\Lambda$ [35] and the phase space, respectively:

$$\rho_n(s_{ik}) = \left(\alpha(n)\frac{s_{ik}}{(m_i+m_k)^2} + \beta(n) + \delta(n)\frac{(m_i-m_k)^2}{s_{ik}}\right) \times$$
$$\times \frac{\sqrt{[s_{ik}-(m_i+m_k)^2][s_{ik}-(m_i-m_k)^2]}}{s_{ik}},\quad (7)$$

The coefficients $\alpha(n)$, $\beta(n)$ and $\delta(n)$ are given in Table IV.

Here n corresponds to a $q\bar{q}$ pair in the color state $1_c$. In the case in question, the interacting quarks do not produce a bound state, therefore the integration in Eqs. (8) - (15) is carried out from the threshold $(m_i+m_k)^2$ to the cutoff $\Lambda(ik)$. The coupled integral equation systems, corresponding to Fig. 1 (the meson state with n=5 and $J^P = 2^-$ for the $c\bar{s}u\bar{u}$) can be described as:

$$A_1(s,s_{12},s_{34}) = \frac{\lambda_1 B_4(s_{12})B_1(s_{34})}{[1-B_4(s_{12})][1-B_1(s_{34})]} + 2\hat{J}_2(s_{12},s_{34},4,1)A_5(s,s'_{23},s'_{123}) +$$
$$+ 2\hat{J}_2(s_{12},s_{34},4,1)A_6(s,s'_{14},s'_{134}) \quad (8)$$

$$A_2(s,s_{23},s_{14}) = \frac{\lambda_2 B_1(s_{23})B_4(s_{14})}{[1-B_1(s_{23})][1-B_4(s_{14})]} + 2\hat{J}_2(s_{23},s_{14},1,4)A_7(s,s'_{34},s'_{234}) +$$
$$+ 2\hat{J}_2(s_{23},s_{14},1,4)A_8(s,s'_{12},s'_{123}) \quad (9)$$

$$A_3(s,s_{12},s_{34}) = \frac{\lambda_3 B_1(s_{12})B_4(s_{34})}{[1-B_1(s_{12})][1-B_4(s_{34})]} + 2\hat{J}_2(s_{12},s_{34},1,4)A_5(s,s'_{23},s'_{123}) +$$
$$+ 2\hat{J}_2(s_{12},s_{34},1,4)A_6(s,s'_{14},s'_{134}) \quad (10)$$

$$A_4(s,s_{23},s_{14}) = \frac{\lambda_4 B_4(s_{23})B_1(s_{14})}{[1-B_4(s_{23})][1-B_1(s_{14})]} + 2\hat{J}_2(s_{23},s_{14},4,1)A_7(s,s'_{34},s'_{234}) +$$
$$+ 2\hat{J}_2(s_{23},s_{14},4,1)A_8(s,s'_{12},s'_{123}) \quad (11)$$

$$A_5(s,s_{23},s_{123}) = \frac{\lambda_5 B_5(s_{23})}{1-B_5(s_{23})} + 2\hat{J}_3(s_{23},5)A_1(s,s'_{12},s'_{34}) + 2\hat{J}_3(s_{23},5)A_3(s,s'_{12},s'_{34}) +$$
$$+ \hat{J}_1(s_{23},5)A_7(s,s'_{34},s'_{234}) + \hat{J}_1(s_{23},5)A_8(s,s'_{12},s'_{123}) \quad (12)$$

$$A_6(s,s_{14},s_{134}) = \frac{\lambda_6 B_5(s_{14})}{1-B_5(s_{14})} + 2\hat{J}_3(s_{14},5)A_1(s,s'_{12},s'_{34}) + 2\hat{J}_3(s_{14},5)A_3(s,s'_{12},s'_{34}) +$$
$$+ \hat{J}_1(s_{14},5)A_7(s,s'_{34},s'_{234}) + \hat{J}_1(s_{14},5)A_8(s,s'_{12},s'_{123}) \quad (13)$$

$$A_7(s,s_{34},s_{234}) = \frac{\lambda_7 B_5(s_{34})}{1-B_5(s_{34})} + 2\widehat{J}_3(s_{34},5)A_4(s,s'_{23},s'_{14}) + 2\widehat{J}_3(s_{34},5)A_2(s,s'_{23},s'_{14}) +$$
$$+ \widehat{J}_1(s_{34},5)A_5(s,s'_{23},s'_{123}) + \widehat{J}_1(s_{34},5)A_6(s,s'_{14},s'_{134})$$
(14)

$$A_8(s,s_{12},s_{123}) = \frac{\lambda_8 B_5(s_{12})}{1-B_5(s_{12})} + 2\widehat{J}_3(s_{12},5)A_4(s,s'_{23},s'_{14}) + 2\widehat{J}_3(s_{12},5)A_2(s,s'_{23},s'_{14}) +$$
$$+ \widehat{J}_1(s_{12},5)A_6(s,s'_{14},s'_{134}) + \widehat{J}_1(s_{12},5)A_5(s,s'_{23},s'_{123})$$
(15)

where $\lambda_i$ are the current constants. They do not affect the mass spectrum of tetraquarks. We introduce the integral operators:

$$\widehat{J}_1(s_{12},l) = \frac{G_l(s_{12})}{[1-B_l(s_{12})]} \int_{(m_1+m_2)^2}^{(m_1+m_2)^2 \Lambda/4} \frac{ds'_{12}}{\pi} \frac{G_l(s'_{12})\rho_l(s'_{12})}{s'_{12}-s_{12}} \int_{-1}^{+1} \frac{dz_1}{2},$$
(16)

$$\widehat{J}_2(s_{12},s_{34},l,p) = \frac{G_l(s_{12})G_p(s_{34})}{[1-B_l(s_{12})][1-B_p(s_{34})]} \times$$
$$\times \int_{(m_1+m_2)^2}^{(m_1+m_2)^2 \Lambda/4} \frac{ds'_{12}}{\pi} \frac{G_l(s'_{12})\rho_l(s'_{12})}{s'_{12}-s_{12}} \int_{(m_3+m_4)^2}^{(m_3+m_4)^2 \Lambda/4} \frac{ds'_{34}}{\pi} \frac{G_p(s'_{34})\rho_p(s'_{34})}{s'_{34}-s_{34}} \int_{-1}^{+1} \frac{dz_3}{2} \int_{-1}^{+1} \frac{dz_4}{2},$$
(17)

$$\widehat{J}_3(s_{12},l) = \frac{G_l(s_{12},\widetilde{\Lambda})}{1-B_l(s_{12},\widetilde{\Lambda})} \times$$
$$\times \frac{1}{4\pi} \int_{(m_1+m_2)^2}^{(m_1+m_2)^2 \widetilde{\Lambda}/4} \frac{ds'_{12}}{\pi} \frac{G_l(s'_{12},\widetilde{\Lambda})\rho_l(s'_{12})}{s'_{12}-s_{12}} \int_{-1}^{+1} \frac{dz_1}{2} \int_{-1}^{+1} dz \int_{z_2^-}^{z_2^+} dz_2 \frac{1}{\sqrt{1-z^2-z_1^2-z_2^2+2zz_1z_2}},$$
(18)

where $l,p$ are equal to 1 - 5. Here $m_i$ is a quark mass.

In Eqs. (16) and (18) $z_1$ is the cosine of the angle between the relative momentum of the particles 1 and 2 in the intermediate state and the momentum of the particle 3 in the final state, taken in the c.m. of particles 1 and 2. In Eq. (18) $z$ is the cosine of the angle between the momenta of the particles 3 and 4 in the final state, taken in the c.m. of particles 1 and 2. $z_2$ is the cosine of the angle between the relative momentum of particles 1 and 2 in the intermediate state and the momentum of the particle 4 in the final state, is taken in the c.m. of particles 1 and 2. In Eq. (17): $z_3$ is the cosine of the angle between relative momentum of particles 1 and 2 in the intermediate state and the relative momentum of particles 3 and 4 in the intermediate state, taken in the c.m. of particles 1 and 2. $z_4$ is the cosine of the angle between the relative momentum of the particles 3 and 4 in the intermediate state and that of the momentum of the particle 1 in the intermediate state, taken in the c.m. of particles 3, 4.

We can pass from the integration over the cosines of the angles to the integration over the subenergies [36].

Let us extract two-particle singularities in the amplitudes $A_1(s, s_{12}, s_{34})$, $A_2(s, s_{23}, s_{14})$, $A_3(s, s_{12}, s_{34})$, $A_4(s, s_{23}, s_{14})$, $A_5(s, s_{23}, s_{123})$, $A_6(s, s_{14}, s_{134})$, $A_7(s, s_{34}, s_{234})$, $A_8(s, s_{12}, s_{123})$:

$$A_j(s, s_{ik}, s_{lm}) = \frac{\alpha_j(s, s_{ik}, s_{lm}) B_1(s_{ik}) B_4(s_{lm})}{[1 - B_1(s_{ik})][1 - B_4(s_{lm})]}, \qquad j=1\text{-}4, \qquad (19)$$

$$A_j(s, s_{ik}, s_{ikl}) = \frac{\alpha_j(s, s_{ik}, s_{ikl}) B_5(s_{ik})}{1 - B_5(s_{ik})}, \qquad j=5\text{-}8, \qquad (20)$$

We do not extract three-particle singularities, because they are weaker than two-particle singularities.

We used the classification of singularities, which was proposed in paper [37]. The construction of approximate solution of Eqs. (8) - (15) is based on the extraction of the leading singularities of the amplitudes. The main singularities in $s_{ik} \approx (m_i + m_k)^2$ are from pair rescattering of the particles i and k. First of all there are threshold square-root singularities. Also possible are pole singularities which correspond to the bound states. The diagrams of Fig.1 apart from two-particle singularities have the triangular singularities and the singularities defining the interaction of four particles. Such classification allows us to search the corresponding solution of Eqs. (8) - (15) by taking into account some definite number of leading singularities and neglecting all the weaker ones. We consider the approximation which defines two-particle, triangle and four-particle singularities. The functions $\alpha_1(s, s_{12}, s_{34})$, $\alpha_2(s, s_{23}, s_{14})$, $\alpha_3(s, s_{12}, s_{34})$, $\alpha_4(s, s_{23}, s_{14})$, $\alpha_5(s, s_{23}, s_{123})$, $\alpha_6(s, s_{14}, s_{134})$, $\alpha_7(s, s_{34}, s_{234})$, $\alpha_8(s, s_{12}, s_{123})$ are the smooth functions of $s_{ik}$, $s_{ikl}$, $s$ as compared with the singular part of the amplitudes, hence they can be expanded in a series in the singularity point and only the first term of this series should be employed further. Using this classification, one defines the reduced amplitudes $\alpha_1$, $\alpha_2$, $\alpha_3$, $\alpha_4$, $\alpha_5$, $\alpha_6$, $\alpha_7$, $\alpha_8$ as well as the B-functions in the middle point of the physical region of Dalitz-plot at the point $s_0$:

$$s_0^{ik} = 0.25(m_i + m_k)^2 s_0$$

(21)

$$s_{123} = 0.25 s_0 \sum_{\substack{i,k=1 \\ i \neq k}}^{3} (m_i + m_k)^2 - \sum_{i=1}^{3} m_i^2, \quad s_0 = \frac{s + 2\sum_{i=1}^{4} m_i^2}{0.25 \sum_{\substack{i,k=1 \\ i \neq k}}^{4} (m_i + m_k)^2}$$

Such a choice of point $s_0$ allows us to replace the integral Eqs. (8) - (15) (Fig. 1) by the algebraic equations (22) - (29) respectively:

$$\alpha_1 = \lambda_1 + 2\alpha_5 JB_1(4,1,5) + 2\alpha_6 JB_2(4,1,5),$$ (22)

$$\alpha_2 = \lambda_2 + 2\alpha_7 JB_3(1,4,5) + 2\alpha_8 JB_4(1,4,5),$$ (23)

$$\alpha_3 = \lambda_3 + 2\alpha_5 JB_5(1,4,5) + 2\alpha_6 JB_6(1,4,5),$$ (24)

$$\alpha_4 = \lambda_4 + 2\alpha_7 JB_7(4,1,5) + 2\alpha_8 JB_8(4,1,5),$$ (25)

$$\alpha_5 = \lambda_5 + 2\alpha_1 JC_1(5,4,1) + 2\alpha_3 JC_2(5,1,4) + \alpha_7 JA_1(5) + \alpha_8 JA_2(5),$$ (26)

$$\alpha_6 = \lambda_6 + 2\alpha_1 JC_3(5,4,1) + 2\alpha_3 JC_4(5,1,4) + \alpha_7 JA_3(5) + \alpha_8 JA_4(5),$$ (27)

$$\alpha_7 = \lambda_7 + 2\alpha_4 JC_5(5,4,1) + 2\alpha_2 JC_6(5,1,4) + \alpha_5 JA_5(5) + \alpha_6 JA_6(5),$$ (28)

$$\alpha_8 = \lambda_8 + 2\alpha_4 JC_7(5,4,1) + 2\alpha_2 JC_8(5,1,4) + \alpha_6 JA_7(5) + \alpha_5 JA_8(5),$$ (29)

We use the functions $JA_i(l)$, $JB_i(l,p,r)$, $JC_i(l,p,r)$ ($l,p,r$ = 1 - 5), which are determined by the various $s_0^{ik}$ (Eq. 21). These functions are similar to the functions:

$$JA_4(l) = \frac{G_l^2(s_0^{12}) B_l(s_0^{23})}{B_l(s_0^{12})} \int_{(m_1+m_2)^2}^{(m_1+m_2)^2 \Lambda/4} \frac{ds_{12}'}{\pi} \frac{\rho_l(s_{12}')}{s_{12}' - s_0^{12}} \int_{-1}^{+1} \frac{dz_1}{2} \frac{1}{1 - B_l(s_{23}')},$$ (30)

$$JB_1(l,p,r) = \frac{G_l^2(s_0^{12}) G_p^2(s_0^{34}) B_r(s_0^{23})}{B_l(s_0^{12}) B_p(s_0^{34})} \times$$

$$\times \int_{(m_1+m_2)^2}^{(m_1+m_2)^2 \Lambda/4} \frac{ds_{12}'}{\pi} \frac{\rho_l(s_{12}')}{s_{12}' - s_0^{12}} \int_{(m_3+m_4)^2}^{(m_3+m_4)^2 \Lambda/4} \frac{ds_{34}'}{\pi} \frac{\rho_p(s_{34}')}{s_{34}' - s_0^{34}} \int_{-1}^{+1} \frac{dz_3}{2} \int_{-1}^{+1} \frac{dz_4}{2} \frac{1}{1 - B_r(s_{23}')}$$

(31)

$$JC_3(l,p,r) = \frac{G_l^2(s_0^{12},\widetilde{\Lambda})B_p(s_0^{23})B_r(s_0^{14})}{1-B_l(s_0^{12},\widetilde{\Lambda})} \frac{1-B_l(s_0^{12})}{B_l(s_0^{12})} \times$$

$$\times \frac{1}{4\pi} \int_{(m_1+m_2)^2}^{(m_1+m_2)^2 \widetilde{\Lambda}/4} \frac{ds'_{12}}{\pi} \frac{\rho_l(s'_{12})}{s'_{12}-s_0^{12}} \int_{-1}^{+1} \frac{dz_1}{2} \int_{-1}^{+1} dz \int_{z_2^-}^{z_2^+} dz_2 \frac{1}{\sqrt{1-z^2-z_1^2-z_2^2+2zz_1z_2}} \times \quad (32)$$

$$\times \frac{1}{[1-B_p(s'_{23})][1-B_r(s'_{14})]}$$

$$\widetilde{\Lambda}(ik) = \begin{cases} \Lambda(ik), & \text{if } \Lambda(ik) \leq (\sqrt{s_{123}}+m_3)^2 \\ (\sqrt{s_{123}}+m_3)^2, & \text{if } \Lambda(ik) > (\sqrt{s_{123}}+m_3)^2 \end{cases} \quad (33)$$

The other choices of point $s_0$ do not change essentially the contributions of $\alpha_1$, $\alpha_2$, $\alpha_3$, $\alpha_4$, $\alpha_5$, $\alpha_6$, $\alpha_7$, $\alpha_8$ therefore we omit the indices $s_0^{ik}$. Since the vertex functions depend only slightly on energy it is possible to treat them as constants in our approximation.

The solutions of the system of equations are considered as:

$$\alpha_i(s) = F_i(s,\lambda_i)/D(s), \quad (34)$$

where zeros of $D(s)$ determinants define the masses of bound states of tetraquarks. $F_i(s,\lambda_i)$ determine the contributions of subamplitudes for the tetraquark amplitude.

### III. Calculation results.

The pole of the reduced amplitudes $\alpha_1$, $\alpha_2$, $\alpha_3$, $\alpha_4$, $\alpha_5$, $\alpha_6$, $\alpha_7$, $\alpha_8$ corresponds to the bound state and determines the mass of the meson-meson state with n=5 and $J^P = 2^-$ for the $c\bar{s}u\bar{u}$ (Fig. 1). The quark masses of model $m_{u,d}$=385 MeV, $m_s$=510 MeV coincide with the ordinary masses in our model [18, 19]. In order to fix $m_c$=1586 MeV, the tetraquark mass for the $J^{PC}=2^{++}$ $X(3940)$ is taken into account. The model in question has only two parameters, the cutoff $\Lambda$=10, and the gluon coupling constant $g$=0.794. These parameters are determined by fixing the masses for the $J^{PC}=1^{++}$ $X(3872)$ and $J^{PC}=2^{++}$ $X(3940)$. The masses of meson-meson states with isospin I=0, ½ and the spin-parity $J^P=1^-,2^-$ are predicted (Tables I and II). In our paper, we calculated the mass M=2750 MeV and the width Γ=5 MeV of tetraquark $(c\bar{s})(u\bar{u})$ with the spin-parity $J^P = 2^-$. The similar state $(c\bar{u})(u\bar{u})$ with the spin-parity $J^P = 2^-$ possesses the mass M=2667 MeV and the width about 6 MeV. We considered also the tetraquarks $(c\bar{s})(u\bar{u})$ and $(c\bar{u})(u\bar{u})$ with the spin-parity $J^P = 1^-$. The masses and the widths of these states are equal to M=2696 MeV, $\Gamma$=6 MeV and M=2621 MeV, $\Gamma$=6 MeV respectively. We studied a relativistic four-body problem and investigated the scattering matrix singularities. The meson-meson state consideration with the gluon exchange interaction is similar to the diquark- antidiquark one. The dynamical mixing of the meson-meson states with the four-quark states is considered.

## IV. Conclusion.

In our paper, the four-fermion interaction is an input [Eq. (2)]. The applicability of Eq. (2) is verified by the success of the De Rujula-Georgi-Glashow quark model [34], where only the short-range part of a Breit potential connected with the gluon exchange is responsible for the mass splitting in meson multiplets. The vertex functions are determined by the contribution of the crossing channels. The vertex functions satisfy the Fierz relations. All of these vertex functions are generated from the g constants [Eq. (3)]. The dynamics of quark interactions is defined by the Chew-Mandelstam functions (Table IV). We include only two parameters: the cutoff and gluon coupling constant, which are similar to the paper [38]. The Tables I and II show the contribution of the subamplitudes: $M_1 M_2$, where $M_1$ and $M_2$ correspond to the meson-meson states. The mass distinction of $u$ and $d$ quarks is neglected. Recently, the BaBar Collaboration discovered the new states with the open charm $D(2550)$, $D(2600)$, $D(2750)$ and $D(2760)$ [39].

## Acknowledgments

The work was carried with the support of the Russian Ministry of Education (grant 2.1.1.68.26) and RFBR, Research Project № 13-02-91154.

Table I. Low-lying meson-meson state masses (MeV) and the contributions of subamplitudes to the tetraquark amplitudes (in percent) for the $J^P = 2^-$, n=5.

| $(c\bar{s})(u\bar{u})$ | $J^P = 2^-$ | $(c\bar{u})(u\bar{u})$ | $J^P = 2^-$ | $(c\bar{s})(s\bar{s})$ | $J^P = 2^-$ |
|---|---|---|---|---|---|
| Masses: | 2750 MeV | Masses: | 2667 MeV | Masses: | 2964 MeV |
| $(c\bar{s})_{2^+}(u\bar{u})_{0^-}$ | 16.01 | $(c\bar{u})_{2^+}(u\bar{u})_{0^-}$ | 16.95 | $(c\bar{s})_{2^+}(s\bar{s})_{0^-}$ | 17.78 |
| $(c\bar{u})_{2^+}(u\bar{s})_{0^-}$ | 2.00 | $(c\bar{u})_{0^-}(u\bar{u})_{2^+}$ | 14.55 | $(c\bar{s})_{0^-}(s\bar{s})_{2^+}$ | 16.12 |
| $(c\bar{s})_{0^-}(u\bar{u})_{2^+}$ | 14.41 | | | | |
| $(c\bar{u})_{0^-}(u\bar{s})_{2^+}$ | 1.75 | | | | |

Table II. Low-lying meson-meson state masses (MeV) and the contributions of subamplitudes to the tetraquark amplitudes (in percent) for the $J^{PC} = 1^-$, n=4.

| $(c\bar{s})(u\bar{u})$ | $J^{PC} = 1^-$ | $(c\bar{u})(u\bar{u})$ | $J^{PC} = 1^-$ | $(c\bar{s})(s\bar{s})$ | $J^{PC} = 1^-$ |
|---|---|---|---|---|---|
| Masses: | 2696 MeV | Masses: | 2621 MeV | Masses: | 2900 MeV |
| $(c\bar{s})_{1^+}(u\bar{u})_{0^-}$ | 16.94 | $(c\bar{u})_{1^+}(u\bar{u})_{0^-}$ | 17.71 | $(c\bar{s})_{1^+}(s\bar{s})_{0^-}$ | 18.37 |
| $(c\bar{u})_{1^+}(u\bar{s})_{0^-}$ | 1.73 | $(c\bar{u})_{0^-}(u\bar{u})_{1^+}$ | 15.99 | $(c\bar{s})_{0^-}(s\bar{s})_{1^+}$ | 17.22 |
| $(c\bar{s})_{0^-}(u\bar{u})_{1^+}$ | 15.79 | | | | |
| $(c\bar{u})_{0^-}(u\bar{s})_{1^+}$ | 1.59 | | | | |

Table III. Vertex functions for the table.

| $J^{PC}$ | $G_n^2$ |
|---|---|
| $0^-$ (n=1) | $8g/3 - 4g(m_i + m_k)^2 /(3s_{ik})$ |
| $1^+$ (n=2) | $4g/3$ |
| $2^+$ (n=3) | $4g/3$ |
| $1^-$ (n=4) | $4g/3$ |
| $2^-$ (n=5) | $4g/3$ |

Table IV. Coefficients of Chew-Mandelstam functions for the table.

| $J^{PC}$ | $\alpha$ | $\beta$ | $\delta$ |
|---|---|---|---|
| $0^-$ (n=1) | 1/2 | -e/2 | 0 |
| $1^+$ (n=2) | 1/2 | -e/2 | 0 |
| $2^+$ (n=3) | 3/10 | 1/5-3e/10 | -1/5 |
| $1^-$ (n=4) | 2/3 | -e | 1/3 |
| $2^-$ (n=5) | 1/2 | -e/2 | 0 |

$e = (m_i - m_k)^2 / (m_i + m_k)^2$

Figure captions.

Fig.1. Graphic representation of the equations for the four-quark subamplitudes $A_k$ ($k$=1-8) in the case of n=5 and $J^P = 2^-$ ($c\bar{s}u\bar{u}$).

References.

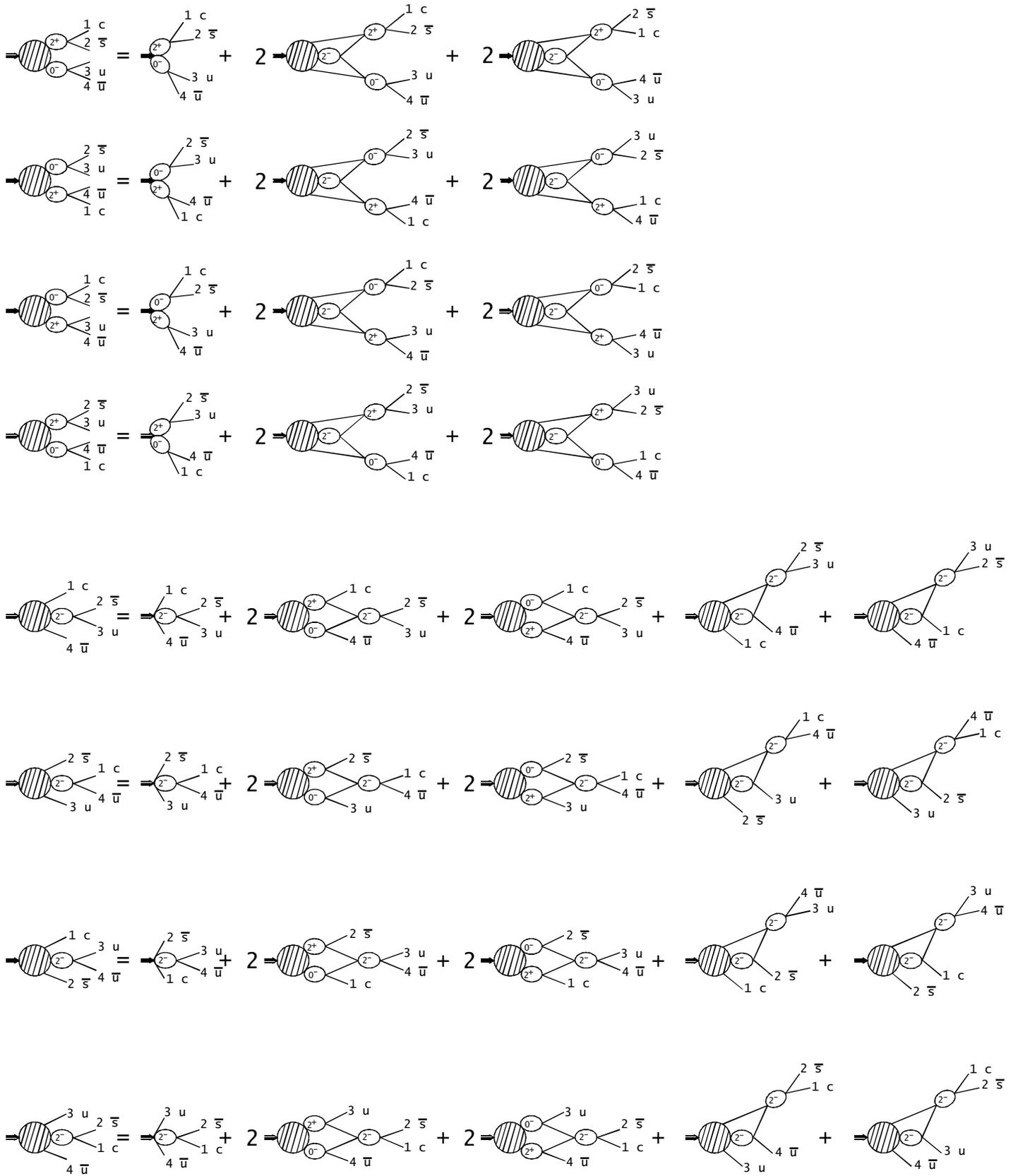

Fig. 1